\documentclass[12pt,titlepage]{article}
\usepackage{amssymb}
\usepackage[dvips]{graphicx}
\begin{document}
\title{On the bound-state spectrum of a nonrelativistic particle in the
background of a short-ranged linear potential}
\author{L.B. Castro\thanks{%
benito@feg.unesp.br } and A.S. de Castro\thanks{%
castro@pq.cnpq.br.} \\
\\
UNESP - Campus de Guaratinguet\'{a}\\
Departamento de F\'{\i}sica e Qu\'{\i}mica\\
12516-410 Guaratinguet\'{a} SP - Brazil}
\date{}
\maketitle

\begin{abstract}
The nonrelativistic problem of a particle immersed in a triangular potential
well, set forth by N.A. Rao and B.A. Kagali, is revised. It is shown that
these researchers misunderstood the full meaning of the potential and
obtained a wrong quantization condition. By exploring the space inversion
symmetry, this work presents the correct solution to this problem with
potential applications in electronics in a simple and transparent way.

\vspace{4in}


\bigskip

\end{abstract}

In a recent paper published in this Journal, Rao and Kagali \cite{rk}
explored the one-dimensional nonrelativistic bound-state solutions of a
particle immersed in a triangular potential well. In view of the mentioned
significance in particle physics and exciting applications in solid state
physics, it is of more than pedagogical interest to revise the problem. The
present paper highlights that the authors of Ref. \cite{rk} misunderstood
the full meaning of the novel potential and made a few erroneous
calculations. Furthermore, the correct spectrum to the triangular potential
well is presented in a simple way.

Let us write the short-ranged linear potential well as
\begin{eqnarray}
V(x) &=&\frac{V_{0}}{L}\left( |x|-L\right) \left[ \theta \left( x+L\right)
-\theta \left( x-L\right) \right]  \nonumber \\
&&  \label{V} \\
&=&\left\{
\begin{array}{c}
\frac{V_{0}}{L}\left( |x|-L\right) \\
\\
0%
\end{array}%
\begin{array}{c}
{\textrm{for }}|x|<L \\
\\
{\textrm{for }}|x|>L%
\end{array}%
\right.  \nonumber
\end{eqnarray}%
where $\theta \left( x\right) $ is the Heaviside function, $2L$ is the range
of the potential and $V_{0}$ is its depth. Because $V\left( -x\right)
=V\left( x\right) $, the Schr\"{o}dinger equation
\begin{equation}
\frac{d^{2}\psi (x)}{dx^{2}}+\frac{2m}{\hslash ^{2}}\left[ E-V\left(
x\right) \right] \psi (x)=0  \label{sch}
\end{equation}%
is invariant under space inversion ($x\rightarrow -x$) and so we can choose
solutions with definite parities. In this circumstance it is enough to
concentrate our attention on one side of the $x$-axis and use the continuity
of $\psi (x)$ and $d\psi (x)/dx$ at the origin, inasmuch as $V(x)$ is
finite. Hence, the two distinct classes of solutions can be discriminated by
the behaviour of $\psi $ and its first derivative at the origin: the
homogeneous Neumann condition at the origin ($d\psi (x)/dx|_{x=0}=0$) for
even parity solutions and the homogeneous Dirichlet condition ($\psi (0)=0$)
for odd ones. We define
\begin{equation}
\varepsilon =\frac{E}{\hslash ^{2}/\left( 2mL^{2}\right) },\quad v_{0}=\frac{%
V_{0}}{\hslash ^{2}/\left( 2mL^{2}\right) }  \label{epsilon}
\end{equation}%
and introduce the new variable%
\begin{equation}
z=\frac{v_{0}^{1/3}}{L}\left[ |x|-L\left( 1+\frac{\varepsilon }{v_{0}}%
\right) \right]  \label{z}
\end{equation}%
so that, for $0<x<L$, the Schr\"{o}dinger equation turns into the Airy
differential equation%
\begin{equation}
\frac{d^{2}\psi (z)}{dz^{2}}-z\psi (z)=0  \label{airy}
\end{equation}%
which has a general solution expressed as a linear superposition of the
linearly independent oscillatory Airy functions $\mathrm{Ai}\left( z\right) $
and $\mathrm{Bi}\left( z\right) $ \cite{abr}%
\begin{equation}
\psi (z)=c_{a}\,\mathrm{Ai}\left( z\right) +c_{b}\,\mathrm{Bi}\left( z\right)
\label{s1}
\end{equation}%
Therefore,
\begin{eqnarray}
&&c_{a}\,\mathrm{Ai}^{\prime }\left( z_{0}\right) +c_{b}\,\mathrm{Bi}%
^{\prime }\left( z_{0}\right) =0\quad \textrm{for even parity solutions}
\nonumber \\
&&  \label{pari} \\
&&c_{a}\,\mathrm{Ai}\left( z_{0}\right) +c_{b}\,\mathrm{Bi}\left(
z_{0}\right) =0\quad \textrm{for odd parity solutions}  \nonumber
\end{eqnarray}%
where $z_{0}$ is the value of $z$ at $x=0$ and the prime means derivative
with respect to $z$. For $x>L$, the evanescent free-particle solution ($\psi
$ must vanish as $x\rightarrow \infty $) is expressed as
\begin{equation}
\psi (x)=c\exp \left( -\frac{\sqrt{-\varepsilon }}{L}x\right)  \label{s2}
\end{equation}%
where $c$ is an arbitrary constant and $\varepsilon <0$. The joining
condition of $\psi $ and its derivative at $x=L$ leads to%
\begin{eqnarray}
c_{a}\,\mathrm{Ai}\left( z_{L}\right) +c_{b}\,\mathrm{Bi}\left( z_{L}\right)
&=&c\,\exp \left( -\sqrt{-\varepsilon }\right)  \nonumber \\
&&  \label{L} \\
c_{a}\,\mathrm{Ai}^{\prime }\left( z_{L}\right) +c_{b}\,\mathrm{Bi}^{\prime
}\left( z_{L}\right) &=&\alpha \,c\,\exp \left( -\sqrt{-\varepsilon }\right)
\nonumber
\end{eqnarray}%
with
\begin{equation}
\alpha =-\frac{\sqrt{-\varepsilon }}{v_{0}^{1/3}}  \label{alpha}
\end{equation}%
and%
\begin{equation}
z_{L}=z_{0}+v_{0}^{1/3}  \label{zL}
\end{equation}%
Combining the top and bottom lines of (\ref{L}) yields%
\begin{equation}
\frac{c_{a}\,\mathrm{Ai}^{\prime }\left( z_{L}\right) +c_{b}\,\mathrm{Bi}%
^{\prime }\left( z_{L}\right) }{c_{a}\,\mathrm{Ai}\left( z_{L}\right)
+c_{b}\,\mathrm{Bi}\left( z_{L}\right) }=\alpha  \label{alpha2}
\end{equation}%
Hence, invoking the segregation between even and odd parity solutions
expressed by (\ref{pari}), one finds
\begin{equation}
\frac{\mathrm{Ai}^{\prime }\left( z_{L}\right) -\alpha \mathrm{Ai}\left(
z_{L}\right) }{\mathrm{Bi}^{\prime }\left( z_{L}\right) -\alpha \mathrm{Bi}%
\left( z_{L}\right) }=\left\{
\begin{array}{c}
\mathrm{Ai}^{\prime }\left( z_{0}\right) /\mathrm{Bi}^{\prime }\left(
z_{0}\right)  \\
\\
\mathrm{Ai}\left( z_{0}\right) /\mathrm{Bi}\left( z_{0}\right)
\end{array}%
\begin{array}{c}
\textrm{for even parity solutions} \\
\\
\textrm{for odd parity solutions}%
\end{array}%
\right.   \label{qc}
\end{equation}%
By solving this quantization conditions one obtains the possible energy
levels by inserting the allowed values of $z_{0}$ in (\ref{z}), i.e.
\begin{equation}
\varepsilon =-v_{0}\left( 1+\frac{z_{0}}{v_{0}^{1/3}}\right)   \label{e1}
\end{equation}%
Hence,%
\begin{equation}
E=-V_{0}\left[ 1+z_{0}\left( \frac{\hslash ^{2}}{2mL^{2}V_{0}}\right) ^{1/3}%
\right]   \label{e2}
\end{equation}%
The numerical computation of $\ z_{0}$ can be done easily with a symbolic
algebra program. The even ($\psi _{+}$) and odd ($\psi _{-}$) parity
eigenfunctions on the entire $x$-axis can be written as%
\begin{eqnarray}
\psi _{\pm }\left( x\right)  &=&\theta \left( +x\right) \left\{ \theta
\left( L-x\right) \left[ c_{a}\,\mathrm{Ai}\left( z\right) +c_{b}\,\mathrm{Bi%
}\left( z\right) \right] +\theta \left( x-L\right) c\,e^{-\sqrt{-\varepsilon
}\,x/L}\right\}   \nonumber \\
&&  \label{fun} \\
&&\pm \theta \left( -x\right) \left\{ \theta \left( x+L\right) \left[ c_{a}\,%
\mathrm{Ai}\left( z\right) +c_{b}\,\mathrm{Bi}\left( z\right) \right]
+\theta \left( -x-L\right) c\,e^{+\sqrt{-\varepsilon }\,x/L}\right\}
\nonumber
\end{eqnarray}%
One can use (\ref{pari}) and the top (or bottom) line of (\ref{L}) to write
the three constants $c_{a}$, $c_{b}$ and $c$ in terms of just one of them.
The remaining constant is to be determinate by normalization.

\begin{figure}[th]
\begin{center}
\includegraphics[width=9cm, angle=0]{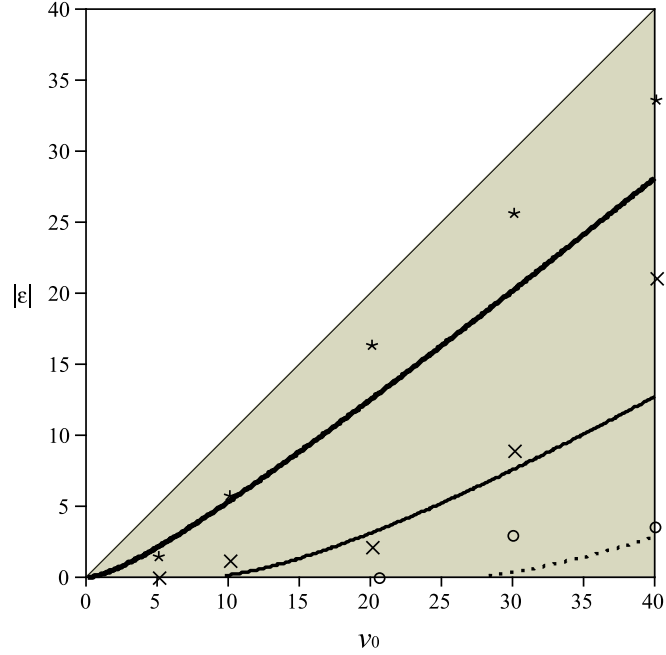}
\end{center}
\par
\vspace*{-0.1cm}
\caption{Absolute values for the ``eigenenergies" ($|\protect\varepsilon |$)
as a function of $v_{0}$ ($L$ \emph{is an arbitrary parameter}). The shaded
area represents the lie zone for bound states ($0<|\protect\varepsilon %
|<v_{0}$). The thick line for the ground state, the thin line for the
first-excited state and the dotted line for the second-excited state. The
asterisks, crosses and circles stand for some values from Table I of Ref.
[1] for the ground, the first-excited and the second-excited states,
respectively.}
\label{Fig1}
\end{figure}

\begin{figure}[th]
\begin{center}
\includegraphics[width=9cm, angle=0]{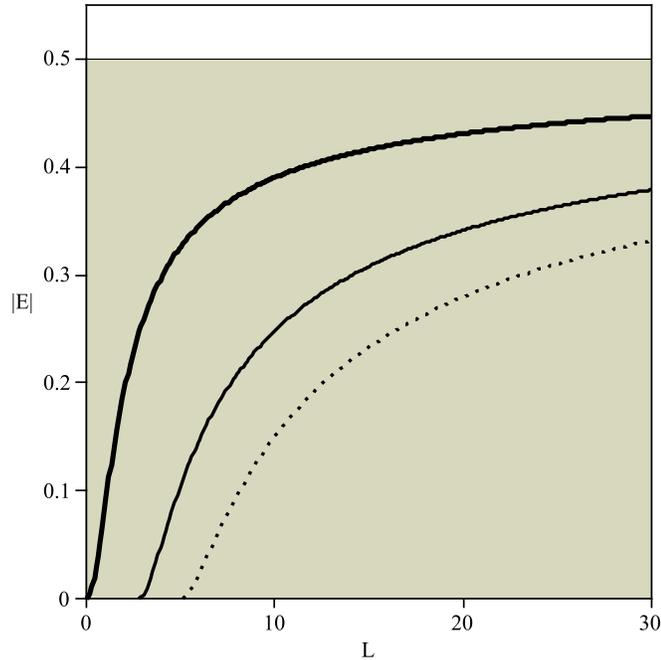}
\end{center}
\par
\vspace*{-0.1cm}
\caption{Absolute values for the eigenenergies ($|E|$) as a function of $L$
for the three lowest states with $V_{0}=0.5$ ($\hslash =m=1$). The shaded
area represents the lie zone for bound states ($0<|E|<V_{0}$). The thick
line for the ground state, the thin line for the first-excited state and the
dotted line for the second-excited state.}
\label{Fig2}
\end{figure}

\begin{figure}[th]
\begin{center}
\includegraphics[width=9cm, angle=0]{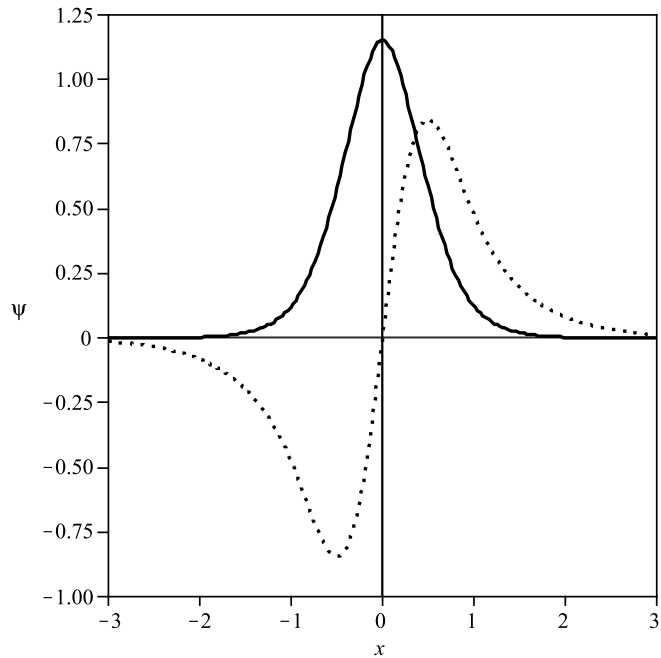}
\end{center}
\par
\vspace*{-0.1cm}
\caption{$\protect\psi $ as a function of $x$ for the ground-state (full
line) and the first-excited state (dotted line), with $L=1$ ($\hslash =m=1$%
), $v_{0}=20$ and $\protect\varepsilon $ equal to $-12.5029801$ and $%
-3.1015082$ respectively.}
\label{Fig3}
\end{figure}

The set of eigenenergies is plotted in Fig. \ref{Fig1} as a function of $%
v_{0}$, and in Fig. \ref{Fig2} as a function of $L.$ The spectra consist of
a finite set of energy levels of alternate parities. Note that the number of
bound states increases with $v_{0}$ and $L$, and that there is always at
least one even parity bound-state solution no matter how weak or narrow the
triangular potential is. Fig. \ref{Fig3} illustrates the behaviour of $\psi
(x)$ for all the states corresponding to $L=1$ and $v_{0}=20$. The
normalization of the eigenfunctions was done by numerical computation using
again a symbolic algebra program.

A peculiar behaviour of the spectrum as $L\rightarrow 0$ can be taken into
account by considering that Airy's functions have the power series expansions \cite{abr}%
\begin{equation}
\mathrm{Ai}\left( z\right) =c_{1}f(z)-c_{2}g(z)\quad \textrm{and}\quad \mathrm{%
Bi}\left( z\right) =\sqrt{3}\left( c_{1}f(z)+c_{2}g(z)\right)  \label{d2}
\end{equation}%
where%
\begin{equation}
f(z)=1+\frac{1}{3!}z^{3}+\frac{4}{6!}z^{6}+\ldots \quad \textrm{and}\quad
g(z)=z+\frac{2}{4!}z^{4}+\ldots  \label{d3}
\end{equation}%
with%
\begin{eqnarray}
c_{1} &=&\mathrm{Ai}\left( 0\right) =\mathrm{Bi}\left( 0\right) /\sqrt{3}%
=3^{-2/3}/\Gamma \left( 2/3\right)  \nonumber \\
&&  \label{d4} \\
c_{2} &=&-\mathrm{Ai}^{\prime }\left( 0\right) =\mathrm{Bi}^{\prime }\left(
0\right) /\sqrt{3}=3^{-1/3}/\Gamma \left( 1/3\right)  \nonumber
\end{eqnarray}%
To be specific, let us look at the case $V_{0}=\lambda /L$, where $\lambda $
is a positive constant. Then
\begin{equation}
z_{0}\sim L^{1/3},\quad z_{L}\sim L^{1/3},\quad \alpha \sim L^{2/3},\quad
v_{0}\sim L,\quad \varepsilon \sim L^{2}  \label{d1}
\end{equation}%
when $L$ is taken to be a small number. When the series (\ref{d2}) are
inserted in (\ref{qc}) and the like powers of $L$ are collected one sees
that the triangular potential does not acquiesce odd parity solutions for
very small $L$. This can be concluded even in the lowest order.
Nevertheless, the four-order approximation in $z$ furnishes%
\begin{equation}
z_{L}^{2}=z_{0}^{2}+2\alpha  \label{d5}
\end{equation}%
for even parity solutions, which combined with (\ref{zL}) gives%
\begin{equation}
\varepsilon =-\frac{v_{0}^{2}}{4}
\end{equation}%
and%
\begin{equation}
\psi (x)=c\exp \left( -\frac{v_{0}}{2L}|x|\right)
\end{equation}%
That is to say, the triangular potential only supports one bound-state
solution. Of course! After all, the triangular potential goes over to the
Dirac delta potential as $L\rightarrow 0$, that is $V(x)\rightarrow -\lambda
\delta (x)$.

Comparison of our results (see Fig. 1) with Table I in Ref. \cite{rk} shows
that the results fail to agree. The reason for this disagreement are a few
mistakes in Ref. \cite{rk}. In Eqs. (5), (6) and (7) of Ref. \cite{rk} the
authors should consider $|y|$ instead of $y$ in the first change of
variable. That \textit{quid pro quo} propagates the error to the continuity
conditions at the origin and makes the quantization condition wrong and too
intricate.

A word should be said about the potential significance of the triangular
well as a quark confining model. The short-ranged linear potential admits
both bound states ($-v_{0}<\varepsilon <0$) and scattering states ($%
\varepsilon >0$). Therefore, it is not a confining potential even though it
is a binding one. A true confining potential, as one of those ones used in
the phenomenological description of the quarkonium, should go to infinity as
$|x|\rightarrow \infty $, even in a relativistic scheme.

Despite the pointed out drawbacks, the authors of the present work recognize
that Rao and Kagali are high-spirited in pursuing such a simple problem
never done before. A meritorious research apart from its potential
applications in electronics. Of course, the investigation of the
nonrelativistic scattering states as well as the extension to the
relativistic domain are worthy.

\bigskip

\bigskip

\noindent \textbf{Acknowledgments}

\noindent This work was supported in part by means of funds provided by
CAPES and CNPq.

\end{document}